# Reflective Fiber Faults Detection and Characterization Using Long-Short-Term Memory

Khouloud Abdelli,[1,3] * Helmut Grießer,[1] Peter Ehrle,[2] Carsten Tropschug,[2] Stephan Pachnicke[3]

[1]ADVA Optical Networking SE, Fraunhoferstr. 9a, 82152 Munich/Martinsried, Germany
[2]ADVA Optical Networking SE, Märzenquelle 1-3, 98617 Meiningen, Germany
[3] Christian-Albrechts-Universität zu Kiel, Kaiserstr. 2, 24143 Kiel, Germany
*Corresponding author: kabdelli@adva.com

**To reduce operation-and-maintenance expenses (OPEX) and to ensure optical network survivability, optical network operators need to detect and diagnose faults in a timely manner and with high accuracy. With the rapid advancement of telemetry technology and data analysis techniques, data-driven approaches leveraging telemetry data to tackle the fault diagnosis problem have been gaining popularity due to their quick implementation and deployment. In this paper, we propose a novel multi-task learning model based on long short-term memory (LSTM) to detect, locate, and estimate the reflectance of fiber reflective faults (events) including the connectors and the mechanical splices by extracting insights from monitored data obtained by the optical time domain reflectometry (OTDR) principle commonly used for troubleshooting of fiber optic cables or links. The experimental results prove that the proposed method: (i) achieves a good detection capability and high localization accuracy within short measurement time even for low SNR values; and (ii) outperforms conventionally employed techniques.**

## 1. INTRODUCTION

Accurate and fast fault detection and localization in fiber optics is extremely crucial to ensure the optical network's survivability and reliability. Any service outage due to a fiber break results in massive data loss, network disruption, and increase in downtime [1]. As reported by the Federal Communication Commission (FCC), more than one-third of service disruptions are caused by fiber-cable problems [2]. Therefore, it is highly beneficial to monitor and diagnose the fiber links remotely and automatically. The fiber optic monitoring helps optical network operators to plan and schedule their maintenance activities more efficiently and thereby save CAPEX/OPEX and reduce the mean time to repair (MTTR) by quickly discovering and pinpointing the link faults, including fiber misalignment/mismatch, fiber breaks, angular faults, dirt on connectors and macro-bends [3]. This allows to meet service level agreements (SLA) more easily, and to improve the customer satisfaction by minimizing the downtimes and enhancing the network quality.

Fiber optic monitoring has mainly been realized using optical time domain reflectometry (OTDR), a technique based on reflections and Rayleigh backscattering, widely applied for fiber characteristics' measurements and for fiber fault detection and localization [4]. The operating principle of OTDR is similar to that of radar. An OTDR injects optical pulses into the fiber under test. Part of these pulses are reflected or scattered back towards the source as a result of Rayleigh scattering. The amplitude of this reflected signal is recorded as a function of propagation time of the light pulse, which can be converted into the position on the optical fiber [5]. Thus, the recorded OTDR trace (or waveform) can show the positions of the connectors and splices (i.e events) along the fiber. The OTDR traces can be noisy and hard to interpret even by trained and experienced field engineers. The noise might be reduced from the OTDR data by taking the average of multiple OTDR measurements. Although the averaging process can reduce the noise level and thus improve the performance of OTDR event analysis approaches in terms of event detection and fault localization accuracy, it is time consuming. The higher the number of averages, the better the signal-to-noise ratio (SNR) of the OTDR trace becomes, however, the longer it takes to analyze the OTDR data. As in optical networks particularly passive optical network (PON) size increases, the number of the fiber links to be monitored increases and thereby the processing time of the OTDR traces escalates and the complexity of the analysis rises leading to less reliable monitoring [3]. To reduce the complexity of the analysis, a well-known technique is to introduce a reference reflector with strong reflectance at the end of each fiber strand. The integrity of the fiber link is tested by monitoring if the reflection level at the reference position is within a specified range. If not, a repeated OTDR analysis with much higher averaging is carried out to identify and localize potential faults within the fiber. This can be achieved by comparing the OTDR trace to a stored reference trace (baseline trace). Alternatively, it would be beneficial to have a reliable automated diagnostic technique that accurately and rapidly detects and locates events in noisy data obtained by an OTDR instrument without the need to perform a lot of averaging and without requiring trained personnel.

Heretofore, some OTDR event analysis approaches have been proposed. Conventionally, the OTDR event detection technique relies on a two-point method combined with the least square approximation method. Although this technique is simple and easy to implement, it is coarse and noise sensitive. Its event detection capability is not high, and it degrades significantly when analyzing low SNR OTDR traces. Lui et al [6] proposed a connection splice event detection and location method based on Gabor series representation and the Rissanen minimum description length criterion, which roughly estimates the event position. To improve the event localization accuracy, they developed a rank-1 matched subspace algorithm (R1MSDE) based on the subspace detection theory of Scharf and Friedlander for detecting and locating the events [7]. The aforementioned technique R1MSDE requires a large amount of computation and is numerically complex. Other OTDR event detection and localization methods based on wavelet analysis have been investigated, too. The singularities in an OTDR trace are detected in this case by a wavelet transform. These techniques, however, imprecisely detect the location of events for OTDR signals with low SNR. Recently, data-driven fault-detection approaches based on machine learning (ML) approaches have been proposed. Trained on normal and faulty operational or collected data, these models can detect and locate faults. Recurrent Neural Networks and particularly Long Short-Term Memory (LSTM) have been proven to efficiently solve the fault diagnosis problems given sequential trained data. In this respect, we have demonstrated data-driven approaches based on LSTM for laser failure detection and prediction given noisy current sensor data [8-10]. Nyarco-Boateng recently proposed a linear regression model to geographically locate faults in underground optical transmission links, given OTDR data [11].

In this paper, we develop a novel LSTM based approach for detecting and locating reflective faults as well as estimating the reflectance given noisy OTDR data. The architecture of the proposed model consists of an LSTM shared hidden layer distributing the knowledge across multiple tasks (event detection, event localization, and event reflectance) followed by a specific task layer.

The proposed approach is applied to noisy experimental OTDR data, whose SNR values vary from 2 dB to 30 dB.

This study's main contributions are as follows:
- The proposed model effectively improves the reflective fiber fault detection capability for low SNR OTDR traces.
- The proposed approach performs higher detection probability and localization accuracy if the OTDR setup parameters combined with the sequences of OTDR trace are inputs of the model.
- The proposed method outperforms the conventional OTDR data analysis techniques particularly for low SNR values.
- The presented approach detects and locates the reflective events faster. It requires less OTDR measurement time by performing less averaging to achieve high fault detection and localization accuracy compared to conventional techniques.

This paper is organized as follows. Section 2 gives some background information about the multi-task learning concept and the LSTM algorithm. Section 3 describes the experimental setups and the process of data generation and processing. Section 4 presents the architecture of the proposed model and the computational complexity analysis of our approach. The results, showing the performance evaluation of the presented model as well as its comparison with a conventional OTDR approach, are discussed in Section 5. Conclusions are drawn in Section 6.

## 2. BACKGROUND
### 2.1 Multi-Task Learning

Multi-task learning (MLT) is a learning paradigm in ML that aims to improve the generalization performance of multiple tasks by learning them jointly while sharing knowledge across them. It has been widely used in various fields ranging from natural language processing to computer vision. The MLT approaches can be classified as hard- and soft parameter sharing. The hard parameter sharing method is done by sharing the hidden layers with the different tasks (completely sharing the weights and the parameters between all tasks) while preserving task-specific output layers learnt independently by each task. Whereas for the soft parameter sharing approach, a model with its own parameters is learnt for each task and the distance between the parameters of the model is then regularized to encourage similarities among related parameters [12]. In this paper, all the tasks, namely the event detection $T_1$, the position estimation $T_2$ and the reflectance prediction $T_3$, are highly related and can benefit from feature space sharing. Thus, an MLT framework with hard parameter sharing is implemented to learn the tasks simultaneously in order to enhance the generalization capability.

### 2.2 Long-short Term Memory

LSTM, originally proposed by Hochreiter and Schmidhuber in 1997 as a solution to the gradient vanishing problem [13], is a special kind of a Recurrent Neural Network (RNN) used to handle a long time series or sequence data achieving state-of-the-art performance in many sequence classification problems such as speech recognition or natural language processing.

The core computational unit of LSTM is called memory cell or block memory (or just cell), consists of weights and three gates, governing the flow information to the cell state, namely the input, the forget and the output gate. The gates are using sigmoid functions having the same equation with different input parameters.

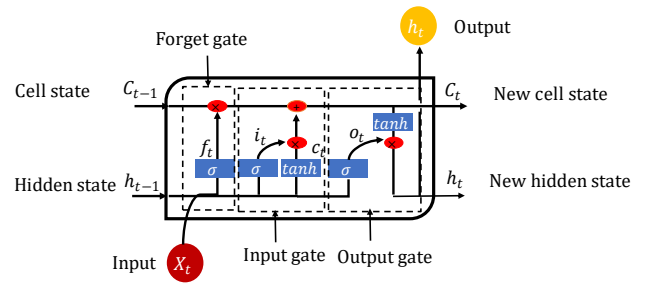

Fig. 1: Structure of a Long Short-Term Memory (LSTM) cell [14].

The equations below describe the functionality of LSTM gates shown in Fig. 1.

$$\begin{aligned}
f_t &= \sigma(x_t \cdot U^f + h_{t-1} \cdot W^f) \\
i_t &= \sigma(x_t \cdot U^i + h_{t-1} \cdot W^i) \\
\tilde{C}_t &= \tanh(x_t \cdot U^g + h_{t-1} \cdot W^g) \\
C_t &= \sigma(f_t \cdot C_{t-1} + i_t \cdot \tilde{C}_t) \\
o_t &= \sigma(x_t \cdot U^o + h_{t-1} \cdot W^o) \\
h_t &= o_t \cdot \tanh(C_t)
\end{aligned} \quad (1)$$

where $\sigma$ is the logistic sigmoid function, $f_t$ the forget gate, $i_t$ the input gate, $o_t$ the output gate, $U$ the weights' matrix, $W$ the recurrent connection at the previous hidden layer and the current layer and $h_{t-1}$ the previous hidden layer. $C$ is a "candidate" hidden state computed based on the current input and the previous hidden states. $C_{t-1}$ is the internal memory of the unit, a combination of the previous memory, multiplied by the forget gate, and the newly computed hidden state, multiplied by the input gate [14].

## 3. SETUP & CONFIGURATION

### 3.1 Proposed Approach Workflow

Figure 2 shows the workflow of the proposed ML based approach for fiber reflective event detection and characterization presented in this work. As illustrated in Fig.2, our approach allows: (i) to detect the faulty fiber links monitored by placing a reflector at the end of each fiber, more quickly; and (ii) to detect, locate and estimate the reflectance of the connectors and the mechanical splices (reflective events) in the fiber links. Note that in operation OTDR trace would be segmented into fixed length sliding windows and our model would be then applied to each segmented sequence to detect and characterize the reflective events. As the position of the reflector is known, just the sequence including the reflective peak is fed to the ML method to investigate the integrity of the link. By checking the probability of the reflective peak and the predicted reflectance by our model, the integrity of the link is verified. Once a link is faulty, the ML model would be applied to the remaining OTDR segmented sequences to detect, locate the reflective faults and to predict the reflectivity of events.

By learning the pattern of the reflective event and by selecting the length of the sequence to be fed to the ML model as small (the probability of existence of two reflective events within the sequence is low), our model could be able to detect the different separate reflective events within the OTDR traces.

### 3.2 Single Event Experimental Setup

The experimental setup shown in Fig. 3 is conducted for recording OTDR traces incorporating reflective faults with small and large peaks. The reflective event is induced by placing a reflector at the end of the fibers under test [15]. The reflector is a wavelength-selective reflector with 95% reflectance that reflects only the test light pulses at 1650 nm sent by OTDR while the wavelength of data signals can pass with small attenuation. The selected experimental setup parameters (small pulse width, low power…) lead to reduce the dynamic range and thereby to weaken the signal in front of the reflective event by increasing the noise relative to the reflected pulse. Given that the signal in front of the reflective peak is too low due to the selected parameters, changing the height of the reflective event in that case would have a similar effect as modifying the reflectance of the reflector. Therefore, we modified the height of the peak by using an internal variable optical attenuator (VOA) in front of an avalanche photodiode (APD) in the OTDR device, in order to generate reflective events with small peaks.

The OTDR instrument sends pulses (probes) at 1650 nm that are additionally attenuated by 13.5 dB using external VOA before being launched into the fibers. By reaching the reflector, the signals are reflected back to the OTDR. Before being received by APD in the OTDR device, we additionally attenuated the signals using an internal VOA with attenuation between 0 to 20 dB. As the averaging of the OTDR measurements influences the SNR of the OTDR trace, 62 up to 64,000 OTDR records are collected and averaged. As well we varied the laser power from 0 to 12 dBm to influence the SNR. The OTDR configuration parameters namely the pulse width and the sampling period are set to 50 ns and 8 ns, respectively.

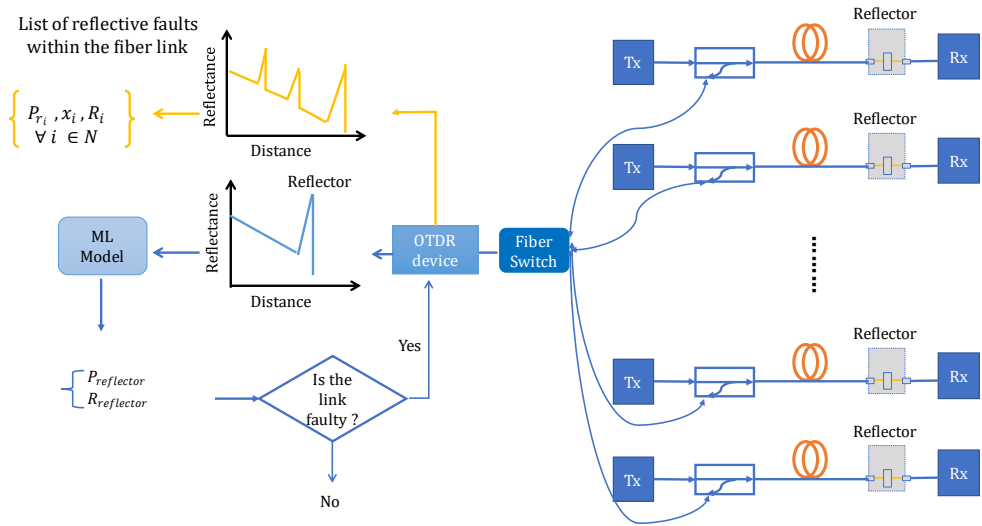

Fig. 2: Typical operation scenario for fiber monitoring with an OTDR device and receiver end reflectors (right) and estimation flow diagram (left) ( $P_{r_i}$: probability of the event, $x_i$: position, $R$: reflectance, $N$: number of events).

Figure 4 depicts an example of an OTDR trace. The first peak at position 0 m is due to the connector between the OTDR and the 750 m fiber. The second peak at 750 m is due to the VOA connectors between the VOA and the first two fibers, the 750 m long fiber and the 5 km fiber. The peak at position 8785 m is the reflective event due to the reflector, which we are looking for (i.e., detecting, locating it and estimating its reflectance). As it can be seen in the trace, the reflective event is hidden in the background noise. This makes detecting and localizing the event challenging. The exponential decay cannot be seen in the trace because of the too low dynamic range due to the selected parameters (the pulse width, the laser power, and the attenuation value of the received signals).

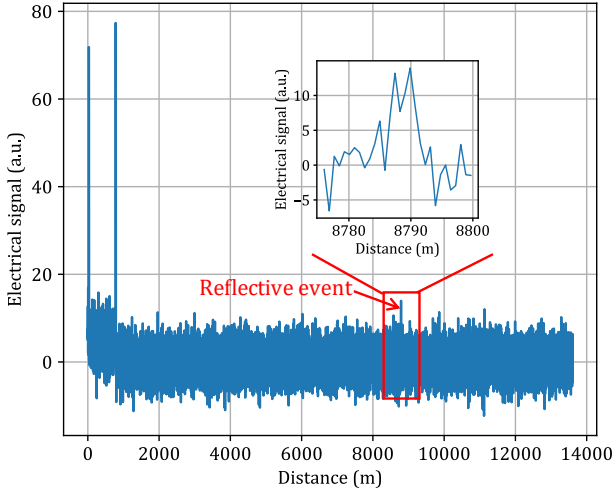

Fig. 4: OTDR trace including the reflective event.

The SNR is defined as

$$SNR = \frac{a}{\sigma_{noise}} \quad (2)$$

where $a$ is the height of the reflective event and $\sigma_{noise}$ is the standard deviation of the noise. The height of the event is estimated as the average of the roof of the peak defined as the average of the two highest values within the range between the event position and the event position plus twice the pulse width length. $\sigma_{noise}$ can be estimated as

$$\sigma_{noise} = \sqrt{\left(\frac{1}{n}\sum_{N-n+1}^{N} x_i^2\right) - \left(\frac{1}{n}\sum_{N-n+1}^{N} x_i\right)^2} \quad (3)$$

where $N$ denotes the last sampling point of the OTDR trace, n is the number of samples used for the estimation (1000 for this data), and $x_i$ is the sampling value.

Up to 6300 traces comprising reflective events with SNR values between 2 and 30 dB and with different reflectance values were generated. Figure 5 shows the distribution of the reflectance. As depicted in Fig. 5, the typical range of reflectance values for the connectors namely Physical Contact (PC) (-40 dB), Superphysical Contact (SPC) (-45 dB), Ultraphysical Contact (UPC) (-50 dB to -55 dB) and Angled Physical Contact (APC) (-60 dB) [16] are included in the histogram modelling the reflectance of the samples used for the training and the test of the ML model. Hence the usefulness and the validity of the data to emulate various real connector scenarios to be learnt by the ML model.

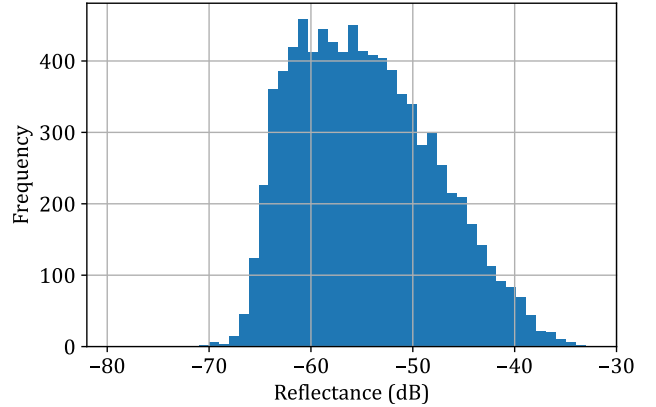

Fig. 5: Reflectance distribution histogram.

### 3.3 Multiple Event Experimental Setup

To evaluate the capability of our model of detecting multiple reflective events with various reflectances within the OTDR trace, the experimental setups shown in Fig. 6 are conducted to generate new unseen test dataset incorporating two reflective events with different reflectances.

The reflector and the open PC connector are used to model the reflective events. The open PC connector has a reflectance of -14 dB, i.e. it reflects 4% of the incoming light. The reflector and the PC connector are interchangeably placed behind the couplers where they are connected to the 5% drop tap. The distance between the reflective faults induced due to reflector or PC connector is varied using 9, 12, 15 or 18 m patch cords. The OTDR configuration parameters are set the same as the ones used in the experimental setup described in Subsection 3.2. The recorded OTDR traces are segmented and fed then to the ML model for testing the detection capability of our approach.

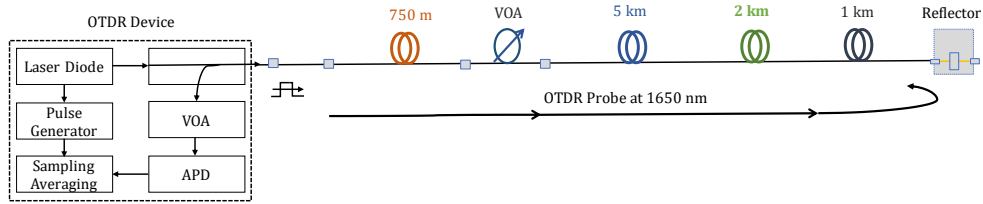

Fig. 3: Experimental setup for a single reflective event (VOA: Variable Optical Attenuator, APD: Avalanche Photodiode).

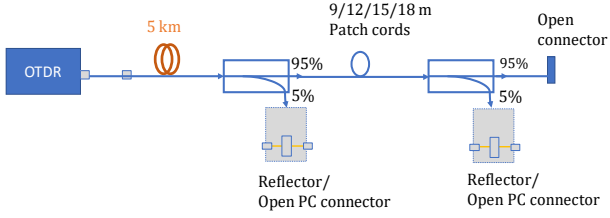

Fig. 6: Multiple reflective event experimental setup.

### 3.4 Data Preprocessing

For training the ML algorithm from each OTDR trace, two sequences of length 35 are extracted randomly: one sequence containing no event, one sequence including the complete reflective event pattern or just part of it in case the reflective peak is on the edge of the window. This ensures that also events that are contained only partially can be detected. In total, a data set composed of 12,600 samples was built. Our proposed model takes a sequence of power levels as input and outputs the $ID_{class}$ (0: no reflective event, 1: reflective event), the reflective event position index within the sequence, and the reflectance R. The reflectance can be derived from the peak height compared to reference measurements for different cleaved fiber ends (see Section VI in [17]).

The generated data (i.e. original data) is normalized and split into a training (60%), a validation (20%) and a test dataset (20%) using random selection.

## 4. MACHINE LEARNING MODEL

In this paper, LSTM is chosen as the shared hidden layer of our MLT framework as it is well suited to extract the time dependency within the OTDR sequential data. The captured time-dependent features (i.e common feature representation/ shared knowledge) are transferred to the task-specific layers to improve the performance of the learning tasks.

### 4.1 Our Proposed Machine Learning Model

The architecture of our model consists of one LSTM hidden layer with 30 neurons, and three layers for each task $T_1$, $T_2$ and $T_3$ composed of 15 neurons. The overall structure of the proposed model is depicted in Fig. 7.

Fed with input sequences consisting of power levels, the LSTM shared layer extracted the useful information underlying the event pattern. The knowledge learned by LSTM is then transferred to the layer of each task. In this way each task layer combines the shared knowledge with the information extracted by itself to improve its generalization performance.

The overall loss function used to update the weights of the model based on the error between the predicted and the desired output can be formulated as

$$L_{total} = \alpha\, L_{T_1} + \beta\, L_{T_2} + \delta\, L_{T_3} \quad (4)$$

where $L_{T_1}$, $L_{T_2}$ and $L_{T_3}$ denote the loss of $T_1, T_2$ and $T_3$, and the first one is the binary cross-entropy loss whereas the others are regression losses (mean squared errors). The loss weights $\alpha, \beta$ and $\delta$ are hyperparameters to be tuned.

### 4.2 Machine Learning Model Computational Complexity

We briefly consider the computational complexity of our proposed model. To approximate the computational time of our model, we have to compute first the complexity of the hidden layer LSTM and that of each task-specific layer. As LSTM is local in space and time (i.e., the input sequence length does not influence its storage requirements), the complexity per weight is $\mathcal{O}(1)$ [13]. Therefore, the computational complexity of LSTM per time step is, where W denotes the number of weights calculated as follows [18]:

$$W = 4 \cdot n_c^2 + 4 \cdot n_{inp} \cdot n_c + n_c \cdot n_{out} + 3 \cdot n_c \quad (5)$$

$n_c$, $n_{inp}$ and $n_{out}$ denote the number of memory cells, the number of input units and the number of output units, respectively.

The computational complexity of task-specific layers can be computed as

$$\mathcal{O}(W_{task}) = j \cdot (n_{out} + n_{task_{out}}) \quad (6)$$

where $j$ denotes the number of nodes in the task specific layer (in our case 15 nodes) and $n_{task_{out}}$ the number of nodes in the output layer of each task. The overall computational complexity of our model is equal to $\mathcal{O}\big(\mathcal{O}(W) + \sum_{i=1}^{3} \mathcal{O}(W_{task_i})\big)$.

The computational complexity of our model is lower than the one of the R1MSDE ($\mathcal{O}(M^3)$) where M denotes the length of the sequence to be processed [7].

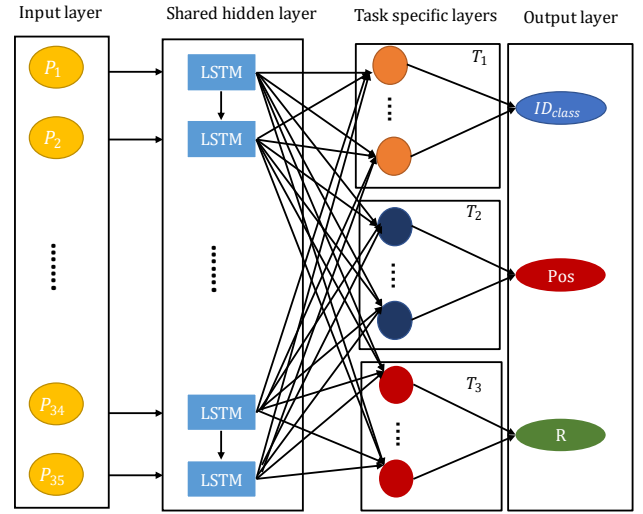

Fig. 7: Proposed model structure.

## 5. RESULTS

### 5.1 Machine Learning Model Overall Performance

We implemented the framework described in the previous section. We tuned the loss weights $\alpha, \beta$ and $\delta$ in order to get better performance. Figure 8 shows the impact of changing the loss weights on the performance of the tasks. In our experiments, we set the loss weights $\alpha, \beta$ and $\delta$ to 0.5, 0.3 and 0.2, respectively since they give better performance as can be seen in Fig.8.

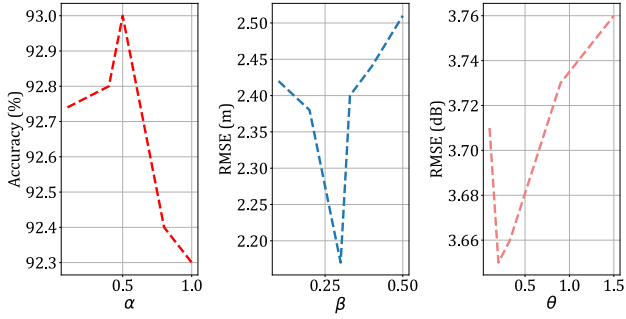

Fig. 8: Effect of changing the loss weights in the task performances.

Figure 9 shows the plot of the learning curve of the ML model during training. As depicted in Fig.9, the performance of the model is good at both the training and the validation data sets. The training and the validation loss slope and the gap between both loss plots is small. That proves that the ML model is not under- or overfitted.

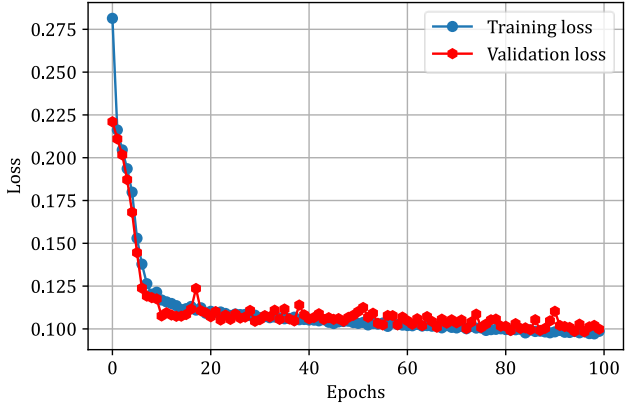

Fig. 9: Learning curves of the ML model.

To evaluate the performance of our proposed model, several metrics were used including the accuracy, the precision quantifying the relevance of the predictions made by the ML model, the recall providing the total relevant results correctly classified by the ML model, and the F1 score, the harmonic mean of the precision and recall, for the event detection task evaluation and the root mean square error (RMSE) and the mean absolute percentage error (MAE) metrics for the evaluation of the event position estimation and reflectance prediction tasks.

The overall performance of our model compared to the single task baselines is depicted in Tab. 1. For the sake of comparison, we used for the single task models the same architecture as our model (one LSTM hidden layer with the same number of hidden neurons, one fully connected layer with the same number of hidden neurons). For the event detection task, the improvement in performance (Δ) over single task model is calculated as the difference between the evaluation metrics of our model and the single task baseline. Whereas for the event position and reflection prediction tasks, Δ is estimated as follows:

$$\Delta = 1 - \frac{m_{multitask}}{m_{single\,task}} \quad \forall\, m \in \{RMSE, MAE\} \quad (7)$$

The experimental results shown in Tab. 1 demonstrate that our model significantly outperforms the single task baselines.

TABLE I
ML MODEL PERFORMANCE EVALUATION

| Metrics | Architecture | | Δ(%) |
|---|---|---|---|
| | Single task | Multitask | |
| **Task 1: Event detection** | | | |
| Accuracy (%) | 92 ±1.06 | 93±0.9 | **+ 1** |
| Precision (%) | 96.5 ±0.7 | 96.6 ±0.7 | **+ 0.1** |
| Recall (%) | 87.2 ±1.3 | 88 ±1.2 | **+ 0.8** |
| F1 score (%) | 91.65±1.1 | 92.1 ±1.05 | **+0.45** |
| **Task 2: Position estimation** | | | |
| RMSE (m) | 4.4 ±0.13 | 2.17±0.07 | **+50.7** |
| MAE (m) | 2.2 ±0.13 | 1.15±0.07 | **+47.7** |
| **Task 3: Reflectance prediction** | | | |
| RMSE (dB) | 5.0± 0.13 | 3.65±0.09 | **+27** |
| MAE (dB) | 3.3 ±0.13 | 2.5±0.09 | **+24.2** |

**5.2 Machine Learning Model Performance as Function of SNR**

Subsequently, we analyzed the performance of our model as function of SNR. Figure 10 shows the effects of SNR on the accuracy of our proposed model. We used the Wilson score [19] to estimate the confidence intervals. The accuracy (A) can be defined as the total number of correctly classified instances divided by the total number of test instances. It is calculated as follows:

$$A = \frac{n_{TP} + n_{TN}}{n_{TP} + n_{TN} + n_{FP} + n_{FN}} \quad (8)$$

where $n_{TP}$, $n_{FN}$, $n_{FP}$, and $n_{TN}$ represent the number of true positives (i.e true reflective event detections), false negatives, false positives, and true negatives (i.e true no reflective event detections), respectively. As expected, the accuracy increases with SNR. For SNR values higher than 5 dB, the accuracy is approaching 1. For an SNR value of 2 dB, the accuracy is worse as it is very difficult to differentiate the event from the noise and thus our model misclassified the event class as normal (i.e. noise). Starting from SNR = 3 dB, our model could significantly distinguish the event from the noise, and it could detect the event with more than 90% accuracy.

Figure 11 presents the RMSE of the event position estimation as function of SNR. When the SNR increases, the RMSE decreases. For lower SNR values (SNR ≤ 10 dB), the RMSE can be higher than 2.5 m, whereas for SNR values higher than 22 dB, it is less than 2 m and it could be further reduced up to less than 1 m for SNR values higher than 30 dB.

Figure 11 also shows that the RMSE of the reflectance prediction for both cases, mixed and whole peak sequences, decreases as SNR increases. For lower SNR values (SNR ≤ 10 dB), RMSE is higher than 3.5 dB, and for higher SNR values, it can be less than 3 dB. Note that the upper blue graph represents the reflectance prediction average error of the instances of reflective events with small and large peaks including the cases of partial reflective event sequences. As indicated by the black graph in Fig.11, the RMSE could be reduced further up to 1.1 dB, if segmentation could ensure that only whole reflective event patterns occur (lower bound).

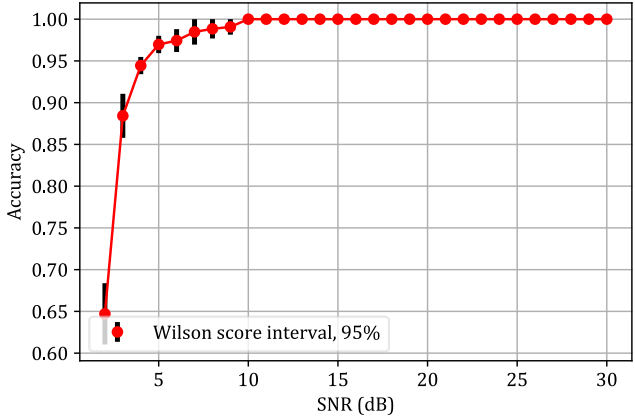

Fig. 10: Event detection performance evaluation with confidence interval.

### 5.3 Experimental Settings for Machine Learning Model Performance

We investigated the impact of including the information about the experimental setup parameters, namely the laser power ($P_{laser}$), the averaging ($N_{avg}$) denoting the number of averaged OTDR traces, the attenuation ($\alpha$) of the received OTDR signals and SNR ($\gamma$) during the training of the ML model to boost the performance. Figure 12 shows the extended architecture of the ML model by adding the experimental setup features as an additional input. First, we trained the model as described in Section III with sequences of signal power levels of length 35. We then added to this sequence the parameters $P_{laser}$, $N_{avg}$ and $\alpha$ influencing the event characteristics, and we trained our model again. We repeated the same process of adding to the initial training data (i.e. sequences of signal power of length 35) the respective parameters $\gamma$, the tuple ($\gamma$, $P_{laser}$, $N_{avg}$, $\alpha$) and the tuple ($\gamma$, $P_{laser}$, $N_{avg}$) and (re-) trained our model.

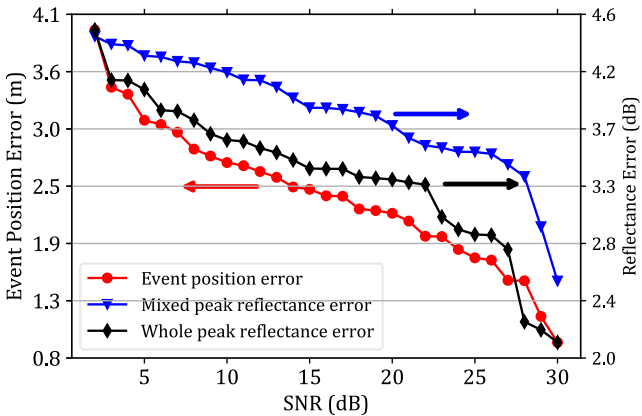

Fig. 11: Event position and reflectance estimation error (RMSE) for the ML model.

The experimental results shown in Fig. 13 demonstrate that adding the experimental settings $P_{laser}$, $N_{avg}$ and $\alpha$ to the training data increases the detection capability of the ML model by 0.75%. Whereas including just $\gamma$ boosts the accuracy by more than 1.5%. This result proves that $\gamma$ is the most important parameter that influences the detection capability of the ML model as the other parameters ($P_{laser}$, $N_{avg}$, $\alpha$) just impact the SNR. Training the ML model with all the experimental setup parameters ($\gamma$, $P_{laser}$, $N_{avg}$, $\alpha$) does not help a lot to enhance the performance of the ML model compared to the case of including $\gamma$ only during the training. This could be explained by the fact that $\gamma$ is highly correlated to $\alpha$. By excluding $\alpha$ and re-training the ML model with the sequence of signal power combined with the remaining settings namely $\gamma$, $P_{laser}$ and $N_{avg}$, the accuracy is improved by more than 2%.

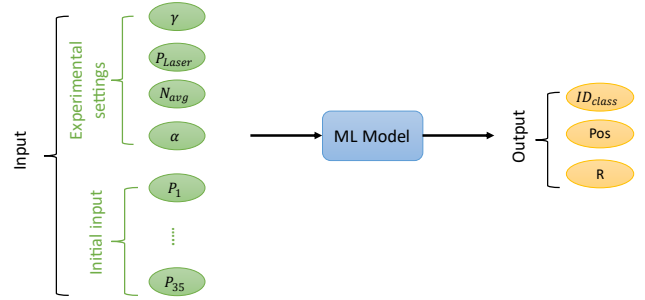

Fig. 12: Extended ML architecture by adding the experimental setup parameters.

The results depicted in Fig. 14 show that including the experimental parameters during the ML model training increases the event localization accuracy. The models trained with experimental settings combined with the initial sequence of power levels could locate the event with lower RMSE up to reaching the resolution limit of 0.2 m for high SNR values.

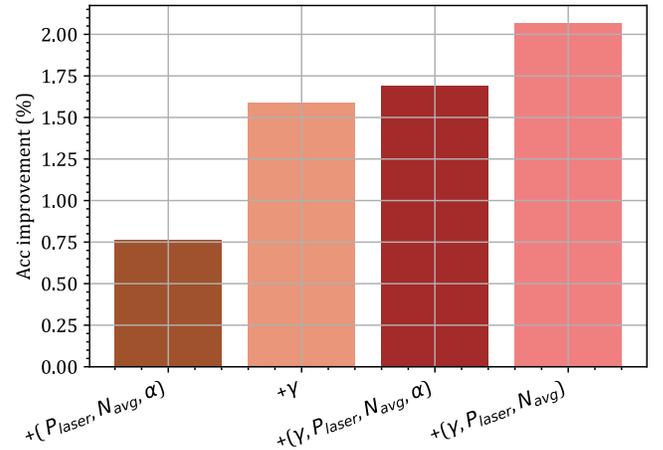

Fig. 13: Accuracy improvement by adding features (experimental settings) to the initial training data (sequences of signal power levels).

As the sampling is 0.8 m, the resolution limit is estimated as $\pm$ 0.4 m but given that the RMSE of the position error is the average value for each SNR, the resolution limit is derived as $\pm$ 0.2 m. For SNR values higher than 8 dB, training the ML model with the sequence of signal power combined with $\gamma$ only gives the best localization accuracy. Including all the experimental setup parameters ($\gamma$, $P_{laser}$, $N_{avg}$, $\alpha$) during the training of the ML model does not improve the event position prediction as the experimental setup parameters are correlated and bring redundant information that does not help the ML model to generalize.

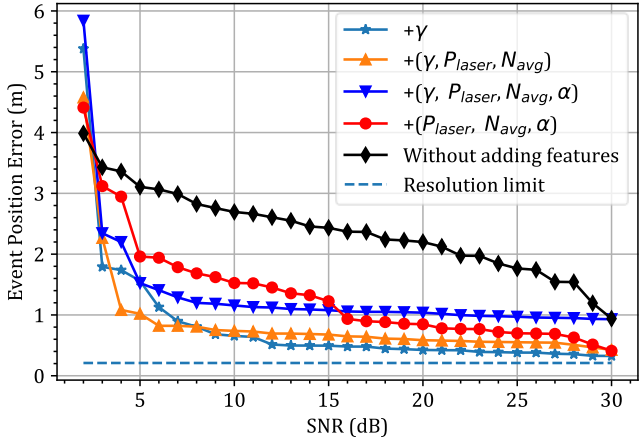

Fig. 14: Event localization accuracy improvement by adding features (experimental settings) to the initial training data (sequences of signal power).

The study of the influence of including the experimental settings during the training of the ML model on the performance prove that training the ML model with $\gamma$ only combined with the sequence of power signal boosts significantly the performance of the ML model in terms of event detection capability and event localization accuracy.

### 5.4 Sequence Length Influence on Machine Learning Model Performance

The performance impact of training the ML model with sequences of different window lengths is explored. We trained the ML model with sequences of length 75, 100, 150 and 200, respectively.

The results shown in Fig. 15 highlight that the ML model's performance decreases with increasing sequence length. It can be explained by the fact that escalating the sequence length leads to increase the number of input features which causes the overfitting of the ML model. For sequence lengths less than 100, the event detection accuracy is higher than 90%, the event position error is less than 7.5 m, and the reflectance prediction error is less than 4 dB. The event position error is strongly affected by the increase of the sequence length. Whereas the reflectance prediction is not highly impacted by training the model with longer sequence.

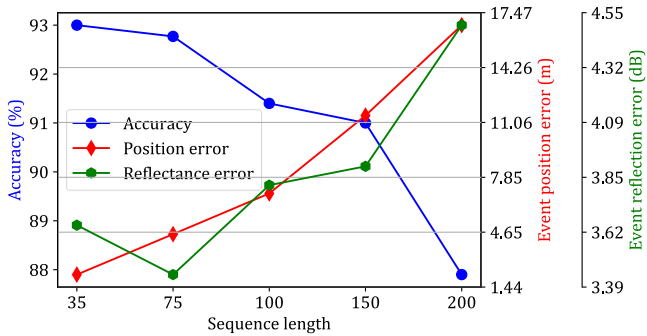

Fig. 15: Sequence length influence on ML model performance.

### 5.5 Machine Learning Model Robustness Investigation

To check the robustness of the ML model, we tested the ML approach on new unseen test data sets slightly different from the data set used to train the ML model. We investigated the following scenarios:

- Estimating the ML performance on test data with a SNR range different from the one of the training data
- Testing the ML model on test data with a reflectance range lower than the one used for the training
- Evaluating the capability of the ML model in detecting multiple reflective events using a test data set incorporating multiple event

#### 5.5.1 Scenarios with Different SNR and Reflectance

We tested the ML model with unseen test data with the SNR range varying from 30 dB to 40 dB. As depicted in Fig.16, our approach is able to detect the event with 99.2% accuracy and locates the event with an RMSE of 1.7 m. The results are expected for such high SNR values, the accuracy of the ML model is close to 100% as it shown in Fig.10.

From the generated two peak data, we extracted just the single reflective event sequences due to the open PC connector with lower reflectance and then we tested the ML model with this data. Figure 16 show that the accuracy of the ML model is 89% and that the model pinpoints the event with a RMSE of 2 m. The results prove that the ML model could generalize well even for the scenarios with low reflectance range.

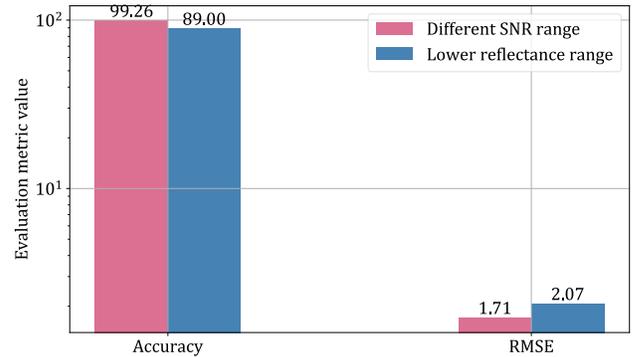

Fig. 16: ML model performance on unseen test data sets with different SNR and reflectance values.

#### 5.5.2 Machine Learning Model Capability of Detecting Multiple Reflective Events

The results of the performance evaluation of the ML model on the unseen two event test data incorporating a mixture of single and double events, depicted in Fig.17, illustrate that our approach detects the reflective faults with 90.2% accuracy and locates them with an RMSE of 1.83 m. Given that the ML model is trained to output if there is an event within the sequence (binary classification), our model can detect the existence of at least one reflective event without discriminating if the detected event is a single or double event. As shown in Fig. 17, the performance of our approach on the two-event data is close to the performance of our model tested on the single event data, discussed in Subsection 4.1. The results demonstrate that our model trained by single event data effectively learnt the pattern of the reflective fault and it is able to generalize well and to detect multiple reflective events within the OTDR trace and even with low reflectances.

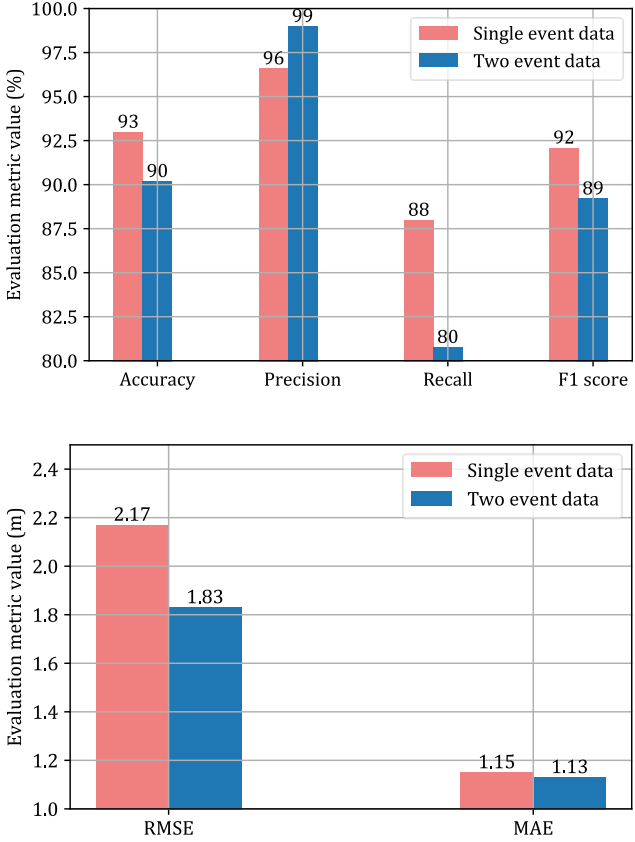

Fig. 17: Performance evaluation of our model on two event data; a) detection-; b) localization evaluation.

### 5.6 Machine Learning Model versus Conventional Method

It should be noted that the developed ML model is trained with data including partial and complete event sequences, since the model should be applicable to arbitrarily segmented OTDR sequences. For a fair comparison of the developed ML model with a conventional rank-1 matched subspace detector (R1MSDE), a test dataset, containing only the complete reflective event pattern or no event was generated. R1MSDE uses the theory of matched subspace detection and associated maximum likelihood estimation procedures to distinguish connection splice events from noise and the Rayleigh component in the OTDR data. As in [7] the reflective event is modelled by a rectangular impulse with prior knowledge of the duration of the event. We optimized the window length of the test data to allow the best performance for R1MSDE. Therefore, we tested the R1MSDE performance with a test dataset composed of sequences of length 100 (which turned out to be the optimum sequence length). Whereas for our model, we evaluated its performance with test sequences of length 35. The detection probability ($P_d$) defined as the portion of the total number of reflective events that were correctly detected, is used to evaluate the detection capability for fixed false alarm probability. It is defined as follows:

$$P_d = \frac{n_{TP}}{n_{TP} + n_{FN}} \quad (9)$$

The false alarm probability ($P_{FA}$) is expressed as:

$$P_{FA} = \frac{n_{FP}}{n_{FP} + n_{TN}} \quad (10)$$

We used the RMSE metric for the evaluation of the event localization accuracy.

A comparison of the different detectors in terms of $P_d$ for $P_{FA}$ of 0.1, as shown in Fig. 18, demonstrates that the ML model outperforms the R1MSDE by achieving higher detection capability particularly for low SNR values.

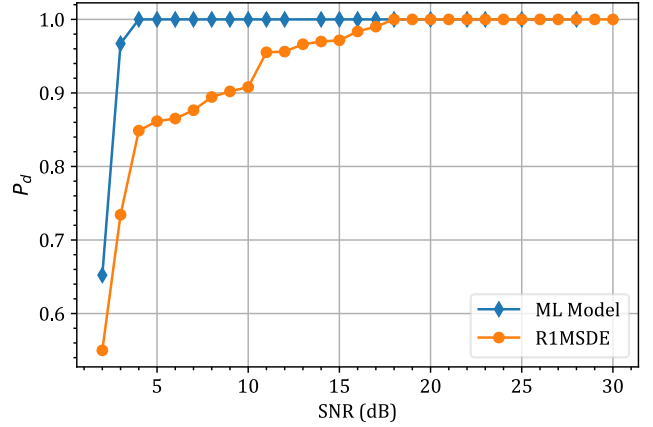

Fig. 18: Comparison of the detection probability for R1MSDE and ML model for a false alarm probability of 0.1.

The results of the comparison of the ML-model's performance with R1MSDE in terms of RMSE of the event position are shown in Fig. 19. For lower SNR values, the ML model outperforms the R1MSDE by achieving a smaller event position error of less than 4 m. As SNR increases, the performance of R1MSDE is getting closer.

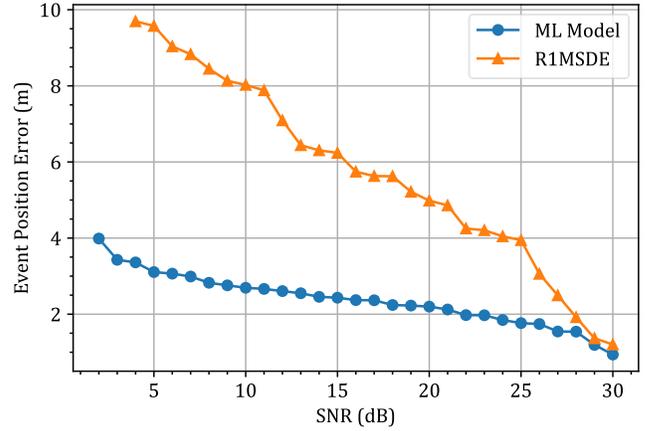

Fig. 19: Event position estimation error (RMSE) for the ML model vs. R1MSDE.

Figure 20 illustrates the results of the comparison of the ML model's accuracy with R1MSDE as function of the averaging time ($\tau$) calculated as follows:

$$\tau\,(s) = \frac{2\,N_{avg}\,L\,n}{c} \quad (11)$$

where $L$ denotes the length of the fibers under test, $n$ is the refractive index and $c$ is the speed of light.

As shown in Fig. 20, our model requires significantly less averaging of OTDR measurements (less measurement time) to achieve the same detection capability as the R1MSDE. This could help to quickly detect the events. To attain an accuracy of 93% using our approach, the averaging time needed is less than 0.2 s whereas for the R1MSDE, it takes more than 12 s.

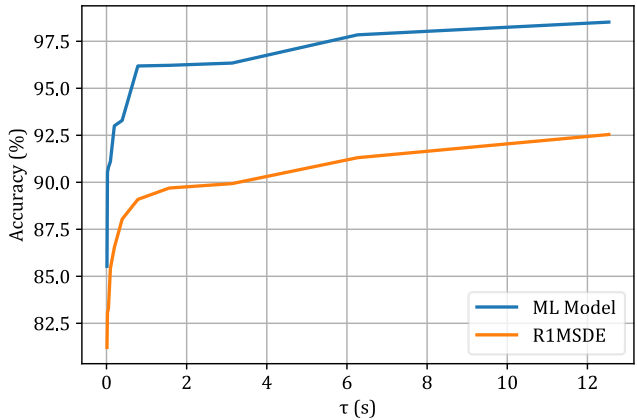

Fig. 20: Comparison of the accuracy for R1MSDE and ML model as function of the averaging time.

## 6. CONCLUSION

In this paper, a multi-task framework based on LSTM for reflective fiber faults' characterization by processing noisy OTDR data is proposed. Trained with a sequence length of 35 of signal power levels, the presented approach learned to detect the reflective event with up to 93% accuracy, to locate it with an RMSE of 2.17 m and to predict its reflectance with an RMSE of 3.65 dB. The experimental results showed that adding the experimental settings and particularly the SNR could significantly enhance the performance of the ML model. The results proved as well that the proposed model outperformed a conventional OTDR event analysis technique. The proposed approach could help to reduce the measurement time required (less averaging of OTDR measurements) to achieve high detection and localization accuracy.

**Acknowledgements.** The work has been partially funded by the German Ministry of Education and Research under project OptiCON (grant No. 16KIS0989K).

## REFERENCES

1. Chan, C.K., F. Tong, L.K. Chen, K.P. Ho, and D. Lim, "Fiber-fault identification for branched access networks using a wavelength-sweeping monitoring source," IEEE Photon. Technol. Lett., 11: 614-616 (1999).
2. A. A.-A. Bakar, M.-Z. Jamaludin, F. Abdullah, M.-H. Yaacob, M. Mahdi, and M. Abdullah, "A new technique of real-time monitoring of fiber optic cable networks transmission," Optics and Lasers in Engineering 45: 126-130 (2007).
3. M.-M. Rad, K. Fouli, H.-A. Fathallah, L.-A. Rusch, and M. Maier, "Passive optical network monitoring: challenges and requirements," IEEE Communications Magazine, vol. 49, no. 2, pp. s45-S52 (2011).
4. Lee W, Myong SI, Lee JC, Lee S. "Identification method of non-reflective faults based on index distribution of optical fibers," Opt Express. 2014 Jan 13;22(1):325-37.
5. Xiaodong Gu and M. Sablatash, "Estimation and detection in OTDR using analyzing wavelets," Proc. of IEEE-SP International Symposium on Time- Frequency and Time-Scale Analysis, pp. 353-356 (1994).
6. Fenglei Liu and C.-J. Zarowski, "Detection and Estimation of Connection Splice Events in Fiber Optics Given Noisy OTDR Data — Part I: GSR / MDL Method," in IEEE Transactions on Instrumentation and Measurement, vol. 50, no. 1, pp. 47-58, (2001).
7. Fenglei Liu and C.-J. Zarowski, "Detection and location of connection splice events in fiber optics given noisy OTDR data. Part II. R1MSDE method," in IEEE Transactions on Instrumentation and Measurement, vol. 53, no. 2, pp. 546-556 (2004).
8. K. Abdelli, D. Rafique, and S. Pachnicke, "Machine Learning Based Laser Failure Mode Detection," ICTON (2019).
9. K. Abdelli, D. Rafique, H. Grießer, and S. Pachnicke, "Lifetime Prediction of 1550 nm DFB Laser using Machine learning Techniques," OFC (2020).
10. K. Abdelli, H. Grießer, and S. Pachnicke, "Machine Learning based Data Driven Diagnostic and Prognostic Approach for Laser Reliability Enhancement," ICTON (2020).
11. O. N-Boateng, A. F-Adekoya, B. A-Weyori, "Predicting the actual location of faults in underground optical networks using linear regression," Engineering Reports (2021).
12. S. Ruder, "An Overview of Multi-Task Learning in Deep Neural Networks," ArXiv abs/1706.05098 (2017).
13. S. Hochreiter and J. Schmidhuber, "Long Short-Term Memory," Neural Computation, 9(8):1735–1780 (1997).
14. S. Varsamopoulos, K. Bertels, and C.G. Almudéver, "Designing neural network based decoders for surface codes," (2018).
15. H. Iida, H. Hirota, T. Uematsu, and K. Noto, "Fresnel Reflection Analysis for Optical Fibre Identification Employing with Three-Wavelength OTDR," ECOC (2020).
16. D.R. Anderson, L. Johnson, F.G. Bell, "Troubleshooting Optical Fiber Networks: Understanding and Using Your Optical Time-Domain Reflectometer" (2004).
17. F.-P. Kapron, B.-P. Adams, E.-A. Thomas, and J.-W. Peters, "Fiber-optic reflection measurements using OCWR and OTDR techniques," IEEE/OSA Journal of Lightwave Technology, vol. 7, no. 8, pp. 1234–1241 (1989).
18. H. Sak, A. Senior, and F. Beaufays, "Long Short-Term Memory Based Recurrent Neural Network Architectures for Large Vocabulary Speech Recognition," ArXiv abs/1402.1128 (2014).
19. Wilson, "Probable inference, the law of succession, and statistical inference", Journal of the American Statistical Association. 22 (158): 209–212, 1927.